\documentclass[english]{article}
\usepackage[T1]{fontenc}
\usepackage[latin9]{inputenc}
\usepackage{float}
\usepackage{textcomp}
\usepackage{amsmath}
\usepackage{amssymb}
\usepackage{graphicx}
\usepackage{esint}
\usepackage{babel}
\usepackage{authblk}
\begin{document}
\title{Adiabatic driving and parallel transport for  parameter-dependent Hamiltonians}
\author[1] {A. D. Berm\'udez Manjarres \thanks{ad.bermudez168@uniandes.edu.co}}
\author[2] {A. Botero \thanks{abotero@uniandes.edu.co}}
\affil[1]{\footnotesize  Universidad Distrital Francisco Jos\'e de Caldas\\ Cra 7 No. 40B-53, Bogot\'a, Colombia\\ }
\affil[2]{\footnotesize Universidad de los Andes\\
Cra. 1E No. 18A-10
Bogot\'a, Colombia\\}
\maketitle
\begin{abstract}
{\normalsize{}We use the Van Vleck-Primas perturbation theory to study
the problem of parallel transport of the eigenvectors of a parameter-dependent Hamiltonian. The perturbative approach allows us to define a non-Abelian connection $\mathcal{A}$ that generates parallel
translation via unitary transformation of the eigenvectors. It is
shown that the connection obtained via the perturbative approach is
an average of the Maurer-Cartan 1-form of the one-parameter subgroup
generated by the Hamiltonian. We use the Yang-Mills curvature and
the non-Abelian Stokes' theorem to show that the holonomy of the connection
$\mathcal{A}$ is related to the Berry phase.}{\normalsize\par}
\end{abstract}

\section{Introduction}

The relation between a parameter-dependent Hamiltonian and its set
of eigenvectors has been studied for a long time in quantum mechanics.
For a small variation of the parameters, this is one of the basic questions
of time-independent perturbation theory. We can also mention the case of a Hamiltonian with slowly varying parameters, where the adiabatic theorem \cite{adiabaticT, adiabaticT2, messiah}, and its generalizations by Berry\cite{Berry} and Wilzeck and Zee \cite{wilzeck}, guarantees that if the system starts in one of the instantaneous eigenstates, and there is no level crossing, then the systems evolve in time very closely as an eigenstate. 

In this work, we review the following problem:
given a family of Hamiltonians $H(\lambda)$ that depend of a set
of external parameters $\lambda$, and assuming we know the set of
eigenvectors of $H(\lambda_{0})$ for some $\lambda_{0}$, how can
we obtain the eigenvectors for any other $H(\lambda)$? 
We will show that the answer involves a parallel transport of the vectors
via a path ordered exponential of a non-Abelian connection 1-form.
Moreover, when the transport around closed loops are considered, the
methods we use will allow us to give a new approach to obtaining the Berry phase \cite{mio}.

The emphasis of this work is the unitary transformations that generate
parallel transport of the basis of eigenvectors of a Hamiltonian $H(\lambda)$
and their relation with the Berry phase. The link between the geometric
phase and unitary transformations has been considered in the past
\cite{U1,U2,U3,U4,U7}. The main difference in the approach we take
here is the use of the perturbation procedure originally developed
by Van Vleck \cite{vv1} and later reformulated in operator form by Primas \cite{vv2}.
The Van Vleck-Primas perturbation theory will allow us to give an operational formula
for the parallel transport operator for finite paths in parameter
space and a new procedure to calculate the Berry phase.

This work is organized as follows: In section 2 we use first-order
van Vleck-Primas perturbation theory to find an infinitesimal unitary
operator $1+i\mathcal{A}(\lambda)\delta\lambda$ that relates the
eigenvectors of Hamiltonian differing by a small change in the external
parameters $\lambda\rightarrow\lambda+\delta\lambda$. The Hermitian operator $\mathcal{A}$  is sometimes called the adiabatic gauge potential, though it is also often used without giving it a proper name. Here, we will call it the adiabatic connection, and our justification is as follows:
first, it will be shown that $\mathcal{A}$  indeed behaves like a non-Abelian
connection 1-form that generates parallel transport. Second, $\mathcal{A}$ has appeared in the literature about the proof of the
adiabatic theorem \cite{adiabaticT, adiabaticT2, messiah} and transitionless quantum driving/shortcuts to adiabaticity \cite{driving, driving2, driving4, driving5,driving6,driving7,driving8,driving9,driving10}, though we point out that our procedures do not depend on the adiabatic theorem or any condition of adiabaticity. In the dynamical case where the parameters
depend on time $\lambda(t)$, $\mathcal{A}$  can be understood as the effective Hamiltonian that produces a transitionless driving along the instantaneous eigenvectors of  $H(\lambda)$.

 We will show that the adiabatic connection can be understood as an
average of the Maurer-Cartan 1-form of the one-parameter subgroup
generated by the instantaneous Hamiltonian.  This results in two formulas to calculate the adiabatic connection (Eqs  (\ref{DefA}) and ( \ref{alternative})).

We point out that the operator $\mathcal{A}$ appears in the context of chaos/integrability, information geometry, topology, and universal properties of phase transitions (see \cite{potential1, potential2, potential4,potential3,potential5,potential6,potential7} and references therein). Hence, this is an important object whose further study is justified.

In section 2.1 we will show how the adiabatic connection defines
a parallel transport of the eigenvectors. In section 2.2 we investigate
the behavior of $\mathcal{A}$ under local gauge transformations,
we prove that it correctly transforms as a non-Abelian connection.

In Section 2.3  we calculate the holonomy of the adiabatic connection, given by a
path ordered exponential $Pe^{\oint_{\gamma}\mathcal{A}}$, and we explicitly show its relation with
the Berry phase. This is a purely geometrical result, as
in no place an adiabatic approximation or even an Schr\"{o}dinger equation
is required. Unlike the usual approaches to the Berry phase, our formula is given in terms of a Yang-Mills curvature 2-form.

In section 3 we apply the results of section 2 to the Hamiltonian
of two well-known systems: a spin interacting with a magnetic field
and a generalized oscillator. In particular, for both examples, the formulas we use for the adiabatic connection allow the calculation to be handled solely using the algebra of the operators involved. In contrast, the usual calculations found in the literature of transitionless driving and time-dependent parameter estimation tend to be performed in terms of the parametrized eigenvectors of the system \cite{driving, driving2, driving4, driving5,driving6,driving7,driving8,driving9,driving10}. 

\section{Adiabatic connection}

Consider a family of Hamiltonians $H(\lambda)$ that depend smoothly
on the set of external parameters $\lambda=\lambda_{1},\lambda_{2}...,\lambda_{N}$
that serve as local coordinates on a manifold $\mathcal{M}$. Suppose
that $R\subset\mathcal{M}$ is a connected region in parameter space
where the Hamiltonian has a discrete, non-degenerate, and possibly
infinite spectrum $\left\{ E_{1},E_{2},\ldots E_{k},\ldots\right\} $.
We assume that the eigenvectors of the Hamiltonian $H(\lambda)$
are known for a given $\lambda\in R$, and we denote them by $\left|n(\lambda)\right\rangle $.
In general, for a small variation in the parameters $\lambda_{\mu}\longrightarrow\lambda_{\mu}+d\lambda_{\mu}$,
the Hamiltonian $H(\lambda+d\lambda)$ will not share the same set
of eigenvectors with $H(\lambda)$. The problem is then to find a
set of eigenvectors $\left|n(\lambda+d\lambda)\right\rangle $ from
the known $\left|n(\lambda)\right\rangle $. To first order, the two
Hamiltonians $H(\lambda)$ and $H(\lambda+d\lambda)$ are related
by

\begin{equation}
H(\lambda+d\lambda)=H(\lambda)+dH(\lambda),
\end{equation}
where $d=d\lambda_{\mu}\frac{\partial}{\partial\lambda_{\mu}}$. Hence,
the problem of finding  $\left|n(\lambda+d\lambda)\right\rangle $
is a problem in time-independent perturbation theory where the perturbation
potential is $dH$. 

A way to find $\left|n(\lambda+d\lambda)\right\rangle $ to first
order is to look for an unitary transformation $e^{i\mathcal{A}(\lambda)}$,
with $\mathcal{A}(\lambda)=A^{\mu}(\lambda)d\lambda_{\mu}$, that
reduces the order of the perturbation from $d\lambda$ to $d\lambda^{2}$
in the following way

\begin{equation}
e^{-i\mathcal{A}}H(\lambda+d\lambda)e^{i\mathcal{A}}=H(\lambda)+D(\lambda)+\mathcal{O}(d\lambda^{2}),\label{vv1}
\end{equation}
where $\mathcal{D}(\lambda)=D_{\mu}(\lambda)d\lambda_{\mu}$ is a
remaining term of order $d\lambda$ that satisfies $[\mathcal{D}(\lambda),H(\lambda)]=0$.
The term $\mathcal{D}(\lambda)$ is known as a shift operator since
the initial energy levels get shifted to 

\begin{equation}
E_{n}(\lambda+d\lambda)=E_{n}(\lambda)+\left\langle n(\lambda)\right|D_{\mu}\left|n(\lambda)\right\rangle d\lambda_{\mu}.\label{shiftE}
\end{equation}

If the operator $\mathcal{A}(\lambda)$ is known, the perturbed eigenvectors
can be written in first order as 

\begin{equation}
\left|n(\lambda+d\lambda)\right\rangle =e^{i\mathcal{A}(\lambda)}\left|n(\lambda)\right\rangle =\left(1+i\mathcal{A}\right)\left|n(\lambda)\right\rangle .\label{iA}
\end{equation}
The procedure to find $e^{i\mathcal{A}(\lambda)}$ is known as the
Van Vleck-Primas perturbation theory \cite{vv1,vv2,vv3,vv4,vv5}.
It can be shown that the problem of finding $\mathcal{A}$ and $\mathcal{D}$
reduces to solve the commutator equations

\begin{eqnarray}
i\left[H,\mathcal{A}\right]+dH & = & \mathcal{D}(\lambda),\label{dif1}\\{}
[\mathcal{D}(\lambda),H(\lambda)] & = & 0.\label{D}
\end{eqnarray}

The above equations have been considered in the literature of shortcuts to adiabaticity, though their relationship with the van Vleck - Primas perturbation theory seems to not be acknowledged. The solution to Eqs (\ref{dif1}) and (\ref{D}) is given in integral
from by \cite{vv2}

\begin{eqnarray}
\mathcal{D}(\lambda) & = & \lim_{T\rightarrow\infty}\int_{0}^{T}ds\,e^{-iHs}\,dH\,e^{iHs}\nonumber \\
 & = & \sum_{n}\left|n(\lambda)\right\rangle \left\langle n(\lambda)\right|dH(\lambda)\left|n,(\lambda)\right\rangle \left\langle n,(\lambda)\right|,
\end{eqnarray}

and

\begin{eqnarray}
\mathcal{A}(\lambda) & =- & \lim_{T\rightarrow\infty}\frac{1}{T}\int_{0}^{T}dt\int_{0}^{t}ds\,e^{-iHs}(dH-\mathcal{D}(\lambda))e^{iHs}\nonumber \\
 & = & i\sum_{n\neq m}\frac{\left|m(\lambda)\right\rangle \left\langle m(\lambda)\right|dH(\lambda)\left|n(\lambda)\right\rangle \left\langle n(\lambda)\right|}{E_{m}-E_{n}}.\label{eq:per}
\end{eqnarray}

Let us note that the operator $\mathcal{A}(\lambda)$ has no diagonal
components in the original base of vectors

\begin{equation}
\left\langle n(\lambda)\right|\mathcal{A}(\lambda)\left|n(\lambda)\right\rangle =0,\;\;\:\forall\left|n(\lambda)\right\rangle .\label{parallelcondition}
\end{equation}

We are solely interested in the variation of the eigenvectors of $H(\lambda)$
and not on its eigenvalues. Therefore, from now on we set $\mathcal{D}=0$
to get $dE_{n}(\lambda)=0$\footnote{This is done without loss of generality because if the condition $dE_{n}(\lambda)=0$
is not met for $H(\lambda)$, we can always define a new family of
Hamiltonians $\widetilde{H}(\lambda)$ with the same eigenvectors
of $H(\lambda)$ but with constant eigenvalues. The unitary transformation
that relates $\left|n(\lambda+d\lambda)\right\rangle $ with $\left|n(\lambda)\right\rangle $
is the same whether we work with $\widetilde{H}(\lambda)$ or with
$H(\lambda)$.}. Thus, the formula for $\mathcal{A}$ reduces to

\begin{equation}
\mathcal{A}(\lambda)=-\lim_{T\rightarrow\infty}\frac{1}{T}\int_{0}^{T}dt\int_{0}^{t}ds\,e^{-isH(\lambda)}(dH)e^{isH(\lambda)}.\label{DefA}
\end{equation}

An alternative integral version for the $\mathcal{A}(\lambda)$ can
be given using the operator identity \cite{dexp1,dexp2}

\begin{equation}
de^{itH}=\int_{0}^{1}ds\,e^{it(1-s)H}(dH)e^{itsH}.\label{identity}
\end{equation}
Inserting Eq (\ref{identity}) into (\ref{DefA}) we get

\begin{equation}
\mathcal{A}(\lambda)=\lim_{T\rightarrow\infty}\frac{i}{2T}\text{\ensuremath{\int}}_{-T}^{T}dt\,e^{-itH}(de^{itH}).\label{alternative}
\end{equation}
We will refer to the operator $\mathcal{A}(\lambda)$ as the adiabatic connection.

Defining the time average of any function of $t$ by

\begin{equation}
\overline{f(t)}=\lim_{T\rightarrow\infty}\frac{1}{2T}\text{\ensuremath{\int}}_{-T}^{T}f(t)\,dt,
\end{equation}
we can write the adiabatic connection (\ref{alternative}) as

\begin{equation}
\mathcal{A}(\lambda)=\overline{\omega(\lambda,t)},\label{averagingA}
\end{equation}
where $\omega(\lambda,t)$ is the left-invariant Maurer-Cartan 1-form of the
one-parameter subgroup generated by the instantaneous $H(\lambda)$
\begin{verse}
\begin{equation}
\omega(\lambda,t)=ie^{-itH(\lambda)}(de^{itH(\lambda)}).\label{maurer-cartan}
\end{equation}
\end{verse}
The adiabatic connection has associated with it a Yang-Mills curvature
2-form 

\begin{equation}
F(\lambda)=\frac{1}{2}F_{\mu\nu}d\lambda_{\mu}\wedge d\lambda_{\nu},\label{curvature2form}
\end{equation}
where the components $F_{\mu\nu}$ are given by

\begin{equation}
F_{\mu\nu}(\lambda)=\partial_{\mu}A_{\upsilon}-\partial_{\upsilon}A_{\mu}-i\left[A_{\mu},A_{\upsilon}\right].\label{curvatureuv}
\end{equation}
The curvature has a simple interpretation in terms of time averages, namely, if we define

\begin{equation}
\delta\omega(t)=\omega(t)-\overline{\omega(t)},
\end{equation}
then 
\begin{equation}
iF_{\mu\nu}=\overline{\left[\delta\omega_{\mu}(t),\delta\omega(t)_{\nu}\right]}.
\end{equation}

To end this section let us note that $\left|n(\lambda+d\lambda)\right\rangle $ can be
expanded as

\begin{equation}
\left|n(\lambda+d\lambda)\right\rangle =\left|n(\lambda)\right\rangle +d\left|n(\lambda)\right\rangle .\label{dn}
\end{equation}
Comparing Eq (\ref{dn}) with Eq (\ref{iA}) we obtain

\begin{equation}
d\left|n(\lambda)\right\rangle =i\mathcal{A}(\lambda)\left|n(\lambda)\right\rangle .\label{parallel Eq}
\end{equation}
From (\ref{parallel Eq}) we can write an alternative formula for the adiabatic connection
\begin{equation}
\mathcal{A}(\lambda)=\sum_{n}\left|dn(\lambda)\right\rangle \left\langle n(\lambda)\right|. \label{AA}
\end{equation}
Equation (\ref{AA}) is the most common formula given for $\mathcal{A}$  in the literature. Comparing  (\ref{AA}) with (\ref{DefA}) and (\ref{alternative}), we see that the latter does not need detailed beforehand knowledge of the variation of the eigenvectors. Moreover,   Eq. (\ref{AA}) requires a sum over all eigenvectors, which can be difficult if the Hamiltonian has an infinite number of eigenvalues. Ultimately, which formula of the adiabatic connection is simpler to work with would depend on the Hamiltonian at hand. In the worked examples given in section 3, the adiabatic connection can be computed using only operator techniques.

We will investigate several of the properties of $\mathcal{A}(\lambda)$ in the next sections.

\subsection{Parallel Transport}

The equation (\ref{parallel Eq}) can be interpreted as a parallel transport
equation where $\mathcal{A}$ takes the role of a non-Abelian connection
(as in general $\left[\mathcal{A}(\lambda),\mathcal{A}(\lambda')\right]\neq0$
for $\lambda\neq\lambda'$). The solution to (\ref{parallel Eq})
is given by a path-ordered exponential

\begin{eqnarray}
\left|n(\lambda),\gamma\right\rangle  & = & \lim_{N\rightarrow\infty}\left(e^{i\mathcal{A}(\lambda_{N})\triangle\lambda_{N}}...e^{i\mathcal{A}(\lambda_{1})\triangle\lambda_{1}}e^{i\mathcal{A}(\lambda_{0})\triangle\lambda_{0}}\right)\left|n(\lambda_{0})\right\rangle \nonumber \\
 & = & Pe^{i\int_{\gamma}\mathcal{A}}\left|n(\lambda_{0})\right\rangle ,\label{path ordered}
\end{eqnarray}
where we have indicated that the final vector depends on the path
in parameter space chosen to go from $\lambda_{0}$ to $\lambda$.

Eq. (\ref{path ordered}) relates the eigenvectors for the Hamiltonian
$H(\lambda_{0})$ with the eigenvectors for $H(\lambda)$ even if
$\lambda$ and $\lambda_{0}$ are not infinitesimally close. To prove
this statement let us note that for $\mathcal{D}=0$ Eq (\ref{dif1})
reduces to

\begin{equation}
dH=i\left[\mathcal{A},H\right].\label{dif3-1}
\end{equation}
Solving Eq (\ref{dif3-1}) for $H$ gives

\begin{equation}
H(\lambda)=Pe^{i\int_{\gamma}\mathcal{A}}H(\lambda_{0})Pe^{-i\int_{\gamma}\mathcal{A}}.
\end{equation}
Therefore,

\begin{equation}
H(\lambda)\left|n(\lambda),\gamma\right\rangle =E_{n}(\lambda_{0})\left|n(\lambda),\gamma\right\rangle .
\end{equation}
The new set of eigenvectors does not only depend on the final point on parameter
space but also on the path $\gamma$ taken to reach it. However, eigenvectors
associated with the same $\lambda$ but reached via different paths
can only differ by a phase factor

\begin{equation}
\left|n(\lambda),\gamma\right\rangle =e^{i\theta}\left|n(\lambda),\gamma'\right\rangle .
\end{equation}

In general, the calculation of an ordered exponential is a complicated
task. However, for closed paths, and under certain circumstances explained
in section 2.3, the path ordered exponential (\ref{path ordered})
can be computed with relative ease.

\subsection{Gauge Transformations}

Let us consider a new family of Hamiltonians,

\begin{equation}
H'(\lambda)=U(\lambda)H(\lambda)U^{\dagger}(\lambda).\label{localgauge}
\end{equation}
The adiabatic connection for the new Hamiltonian is given by

\begin{eqnarray}
\mathcal{A}' & = & i\overline{Ue^{-itH}U^{\dagger}d\left(Ue^{itH}U^{\dagger}\right)}\nonumber \\
 & = & U\mathcal{A}U^{\dagger}+iUdU^{\dagger}+i\overline{Ue^{-itH}\left(U^{\dagger}dU\right)e^{itH}U^{\dagger}}.\label{newA}
\end{eqnarray}
Performing the time average in the last term, we are left with

\begin{eqnarray}
\mathcal{A}' & = & U\mathcal{A}U^{\dagger}+iUdU^{\dagger}+i\sum_{n}\left\langle n(\lambda)\right|\left(U^{\dagger}dU\right)\left|n(\lambda)\right\rangle U\left|n(\lambda)\right\rangle \left\langle n(\lambda)\right|U^{\dagger}\nonumber \\
 & = & U\mathcal{A}U^{\dagger}+i\left(UdU^{\dagger}\right)_{\perp},
\end{eqnarray}
where $\left(UdU^{\dagger}\right)_{\perp}$ is the same as $U^{\dagger}dU$
with the diagonal elements set to zero. The part that is subtracted does not affect the Hamiltonian and only affects the phase of the
eigenvectors, so an equally valid perturbation connection for $H'$
is 

\begin{equation}
\mathcal{A}'=U\mathcal{A}U^{\dagger}+iUdU^{\dagger}.\label{newnewA}
\end{equation}

We can see from Eq (\ref{newnewA}) that $\mathcal{A}$ behaves correctly
as a non-Abelian connection under a local gauge transformation.

\subsection{Holonomy of the Connection}

After a closed circuit in parameter space the vectors obtained by
parallel transport has to be eigenvectors of the original Hamiltonian;
thus, they have to be proportional to the original eigenvectors

\begin{equation}
\left|n(\lambda_{0}),\gamma\right\rangle =e^{i\phi_{n}}\left|n(\lambda_{0})\right\rangle ,\label{originaln}
\end{equation}
where the phase factor in Eq. (\ref{originaln}) will depend on the
path $\gamma$. The holonomy of the adiabatic connection is related
to the Berry phase. To see that this should be the case, let us use
Eq. (\ref{parallel Eq}) to rewrite (\ref{parallelcondition}) as

\begin{equation}
\left\langle n(\lambda)\right|\mathcal{A}(\lambda)\left|n(\lambda)\right\rangle =\left\langle n(\lambda)\right|\left.d\,n(\lambda)\right\rangle =0.\label{berry simon}
\end{equation}
Condition (\ref{berry simon}) is equivalent to the Berry-Simon connection
for the nth vector. Thus, the adiabatic connection should define
a horizontal lift of $\gamma$ with respect to the Berry-Simon connection
for every eigenvector.

The above observation turns out to be correct, after a closed circuit
in parameter space each eigenvector acquires a phase factor that is
equal to the Berry phase. As we are using the adiabatic connection
to generate the parallel transport, it has to be the case that the
geometric phases are encoded in the connection. This statement can be explicitly shown, and we formalize it
in the following lemma:

\paragraph*{Lemma.}

\emph{If $R\subset\mathcal{M}$ is a connected region of parameter
space where $H(\lambda)$ has a non-degenerate spectrum everywhere,
then for any simple closed curve $\gamma$ contained in $R$ it holds
that}

\emph{
\begin{equation}
Pe^{\oint_{\gamma}\mathcal{A}}=diag\,(e^{i\phi_{1}},e^{i\phi_{2}},\ldots,e^{i\phi_{k}},\ldots),\label{lemma}
\end{equation}
where $\phi_{1},\phi_{2},...,\phi_{k},...$ are the Berry phases associated
with the parallel transport around $\gamma$ of $\left|1(\lambda_{0})\right\rangle ,\left|2(\lambda_{0})\right\rangle ,...,\left|k(\lambda_{0})\right\rangle ,....,$
respectively.}
\[
\]

The proof of the above lemma starts with the calculation of the matrix
components of the curvature (\ref{curvatureuv}). We show in Appendix A that the components of Yang-Mills curvature at $\lambda$
are diagonal in the base of vectors $\left|n(\lambda)\right\rangle $,
\begin{eqnarray}
\left\langle n\right|F_{\mu\nu}(\lambda)\left|n\right\rangle  & = & i\sum_{n\neq n'}\left\{ \frac{\left\langle n\right|\partial_{\mu}H\left|n'\right\rangle \left\langle n'\right|\partial_{\nu}H\left|n\right\rangle }{\left(E_{n}-E_{n'}\right)^{2}}-\frac{\left\langle n\right|\partial_{\nu}H\left|n'\right\rangle \left\langle n'\right|\partial_{\mu}H\left|n\right\rangle }{\left(E_{n}-E_{n'}\right)^{2}}\right\} ,\nonumber \\
\left\langle m\right|F_{\mu\nu}(\lambda)\left|n\right\rangle  & = & 0.\label{curv}
\end{eqnarray}
Moreover, the diagonal elements of $F_{\mu\nu}$ are the components
of the Abelian Berry curvature 2-form for the nth vector\cite{Berry}, 

\begin{equation}
\mathcal{W}_{\mu\nu}^{(n)}=\sum_{n\neq n'}\left\{ \frac{\left\langle n\right|\partial_{\mu}H\left|n'\right\rangle \left\langle n'\right|\partial_{\nu}H\left|n\right\rangle }{\left(E_{n}-E_{n'}\right)^{2}}-\frac{\left\langle n\right|\partial_{\nu}H\left|n'\right\rangle \left\langle n'\right|\partial_{\mu}H\left|n\right\rangle }{\left(E_{n}-E_{n'}\right)^{2}}\right\} .\label{Berrycurvature}
\end{equation}
The Berry phase of the nth eigenvector can be calculated from (\ref{Berrycurvature})
as  \cite{Berry, gphase}

\begin{equation}
\phi_{n}=\int_{s}\mathcal{W}^{(n)}.\label{phiw}
\end{equation}

The Yang-Mills curvature is associated with the holonomy of infinitesimal
loops, this is, for sufficiently small $\gamma$ the path ordered
exponential reduces to  \cite{small loop}

\begin{equation}
Pe^{\oint_{\gamma}\mathcal{A}}=e^{i\varepsilon^{2}F(\lambda_{0})},\label{smallF}
\end{equation}
where $\varepsilon^{2}$ is the area element enclosed by $\gamma$.
It follows from (\ref{curv}), (\ref{phiw}) and (\ref{smallF}) that,
for sufficiently small loops, the holonomy of the connection is given
by (\ref{lemma}),

\begin{eqnarray}
\left\langle n(\lambda_{0})\right|Pe^{\oint_{\gamma}\mathcal{A}}\left|m(\lambda_{0})\right\rangle  & = & \delta_{n,m}e^{i\varepsilon^{2}\mathcal{W}^{(n)}(\lambda_{0})}=\delta_{n,m}e^{i\phi_{n}}.\label{infini}
\end{eqnarray}

We can see from Eq. (\ref{infini}) that Eq. (\ref{lemma}) holds
for infinitesimal loops. We will use the result known as the non-Abelian
Stokes' theorem (NAST) \cite{stokes,stokes2,stokes 3,stokes 4} to
extend the proof to finite paths. This theorem roughly states that
the path ordered exponential round a closed curve $\gamma=\partial S$
can be replaced by a surface-ordered exponential

\begin{equation}
Pe^{i\oint_{\gamma}\mathcal{A}}=\mathcal{P}e^{i\int_{S}\mathcal{F}},\label{nonabelian}
\end{equation}
where $\mathcal{F}=U^{-1}FU$ is known as the ``twisted'' curvature,
$U$ is a parallel transport operator and $\mathcal{P}$ represents
a suitable surface ordering. In general, the calculation of the surface
ordered exponential can be harder than the path-ordered one. However,
for the adiabatic connection the computation of the right hand
side of Eq. (\ref{nonabelian}) is straightforward. We will give an
intuitive explanation of the theorem in Appendix B, for a rigorous
proof see \cite{stokes}. For our purpose here it is sufficient to
say that, given parametrization on the surface $(x_{i},x_{j})$, the
right hand side of Eq. (\ref{nonabelian}) can be written as an ordered
product of terms of the form $U_{i,j}^{-1}\,e^{i\varepsilon_{i,j}^{2}F_{i,j}}\,U_{i,j}$
called ``lassos'', where $U_{i,j}$ is a parallel transport operator
labeled by its coordinates on the surface. In terms of the lassos,
the right-hand side of Eq. (\ref{nonabelian}) read

\begin{equation}
\mathcal{P}e^{i\int_{S}\mathcal{F}}=\lim_{N\rightarrow\infty}\mathcal{P}\prod_{i,j}^{N}U_{i,j}^{-1}\,e^{i\varepsilon_{i,j}^{2}F_{i,j}}\,U_{i,j}.\label{PF}
\end{equation}
The exact way that the operators in Eq. (\ref{PF}) have to be ordered
for Eq. (\ref{nonabelian}) to be true will has no importance to us
in what follows. We proceed to prove the lemma by direct calculation
of the matrix elements of $\mathcal{P}e^{i\int_{S}\mathcal{F}}$,
we will evaluate the terms one at a time. Let us focus on the action
of a single lasso at some point $\lambda=\lambda(x_{i},x_{j})$ on
a given eigenvector $\left|n(\lambda_{0})\right\rangle $. First,
the parallel transport operator $U(\lambda,\lambda_{0})$ transforms
a eigenvector at $\lambda_{0}$ to some eigenvector at $\lambda$

\begin{equation}
U(\lambda,\lambda_{0})\left|n(\lambda_{0})\right\rangle =\left|n(\lambda)\right\rangle .
\end{equation}
The next operation is circling the small loop around $\lambda$, where
we get the Berry phase associated with that infinitesimal loop

\begin{equation}
e^{i\varepsilon^{2}F(\lambda)}\left|n(\lambda)\right\rangle =e^{i\varepsilon_{i,j}^{2}\mathcal{W}{}_{i,j}^{(n)}(\lambda)}\left|n(\lambda)\right\rangle .
\end{equation}
Finally, we take back this vector from $\lambda$ to $\lambda_{0}$ 

\begin{equation}
U(\lambda,\lambda_{0})^{-1}e^{i\varepsilon_{i,j}^{2}\mathcal{W}_{i,j}^{(n)}(\lambda)}\left|n(\lambda)\right\rangle =e^{i\varepsilon_{i,j}^{2}\mathcal{W}_{i,j}^{(n)}(\lambda)}\left|n(\lambda_{0})\right\rangle .
\end{equation}
Joining up these three operations we can write

\begin{equation}
U(\lambda,\lambda_{0})^{-1}e^{i\varepsilon^{2}F(\lambda)}U(\lambda,\lambda_{0})\left|n(\lambda_{0})\right\rangle =e^{i\varepsilon^{2}\mathcal{W}^{(n)}(\lambda)}\left|n(\lambda_{0})\right\rangle .\label{intermediate}
\end{equation}

We have that, irrespective of the surface order, all the terms
in the right-hand side of Eq. (\ref{PF}) give rise to phases like
the one in Eq. (\ref{intermediate}). The next result follows from
our previous observation

\begin{eqnarray}
\mathcal{P}e^{i\int_{S}\mathcal{F}}\left|n(\lambda_{0})\right\rangle  & = & \lim_{N\rightarrow\infty}\mathcal{P}\prod_{i,j}^{N}U_{i,j}^{-1}\,e^{i\varepsilon_{i,j}^{2}F_{i,j}}\,U_{i,j}\left|n(\lambda_{0})\right\rangle \nonumber \\
 & = & \lim_{N\rightarrow\infty}\prod_{i,j}^{N}e^{i\varepsilon_{i,j}^{2}\mathcal{W}_{i,j}^{(n)}}\left|n(\lambda_{0})\right\rangle .\nonumber \\
 & = & e^{i\int_{S}\mathcal{W}^{(n)}}\left|n(\lambda_{0})\right\rangle .
\end{eqnarray}
Using the spectral theorem we can finally write

\begin{eqnarray}
Pe^{i\oint_{\gamma}\mathcal{A}} & = & \mathcal{P}e^{i\int_{S}\mathcal{F}}=\sum_{n}e^{i\int_{S}\mathcal{W}^{(n)}}\left|n(\lambda_{0})\right\rangle \left\langle n(\lambda_{0})\right|\nonumber \\
 & = & \sum_{n}e^{i\phi_{n}}\left|n(\lambda_{0})\right\rangle \left\langle n(\lambda_{0})\right|.\blacksquare\label{finalholonomby}
\end{eqnarray}

To end this section we point out that the averaging in the definition
of the adiabatic connection (\ref{averagingA}) is crucial to have
a non-trivial holonomy. The non-averaged 1-form $\omega$ has a vanishing
curvature 2-form due to the Maurer-Cartan equation

\begin{equation}
d\omega+\frac{1}{2}\left[\omega,\omega\right]=0;\label{MCEq}
\end{equation}
hence, a vector transported via $Pe^{i\int_{\gamma}\omega}$ will
return to its original value after a close circuit in parameter space.

\section{Examples}

We calculate the adiabatic connection and its associated curvature
2-form for two well-known Hamiltonians.

\subsection{SU(2)}

Consider the Hamiltonian of a spin-magnetic field interaction given
by

\begin{equation}
H=B\mu(\mathbf{J\cdot\widehat{n}}),
\end{equation}
where the components of $\mathbf{J}=(J_{x},J_{y},J_{z})$ obey the
$SU(2)$ algebra

\begin{equation}
[J_{i},J_{j}]=i\varepsilon_{ijk}J_{k}.
\end{equation}
The parameter to vary in the Hamiltonian is the direction of the
vector $\widehat{\mathbf{n}}$ and the strength of the magnetic field
$B$. 

To avoid degeneracy we restrict our attention to the subspace spanned
by the vectors $\left|l,m\right\rangle _{z}$ with a fixed $l$ and
where

\begin{eqnarray}
J^{2}\left|l,m\right\rangle _{z} & = & l(l+1)\left|l,m\right\rangle _{z},\\
J_{z}\left|l,m\right\rangle _{z} & = & m\left|l,m\right\rangle _{z}.
\end{eqnarray}

The symmetry of the problem suggests the use of spherical coordinates.
Let us make the following definitions

\begin{eqnarray}
J_{n} & = & \mathbf{J}\cdot\widehat{n},\\
J_{\theta} & = & \mathbf{J}\cdot\mathbf{\widehat{\theta}},\\
J_{\phi} & = & \mathbf{J}\cdot\widehat{\mathbf{\phi}}.\label{spherical j}
\end{eqnarray}
The operators just defined give a new representation of the $SU(2)$
algebra

\begin{eqnarray}
[J_{n},J_{\theta}] & = & iJ_{\phi},\\
{}[J_{\phi},J_{n}] & = & iJ_{\theta},\\
{}[J_{\theta},J_{\phi}] & = & iJ_{n}.
\end{eqnarray}
Therefore, there exists a basis of vectors of the form $\left|l,m\right\rangle _{\widehat{n}}$
such that

\begin{eqnarray}
J^{2}\left|l,m\right\rangle _{\widehat{n}} & = & l(l+1)\left|l,m\right\rangle _{\widehat{n}},\\
J_{n}\left|l,m\right\rangle _{\widehat{n}} & = & m\left|l,m\right\rangle _{\widehat{n}}.
\end{eqnarray}

To compute the components of the connection $\mathcal{A}$
we will need the following derivatives

\begin{eqnarray}
\partial_{B}H & = & \mu(\mathbf{J\cdot\widehat{n}}),\\
\partial_{\theta}H & = & B\mu(\mathbf{J\cdot\partial_{\theta}\widehat{n}})=J_{\theta},\\
\partial_{\phi}H & = & B\mu(\mathbf{J\cdot\partial_{\phi}\widehat{n}})=B\mu\sin\theta J_{\phi}.
\end{eqnarray}

The operators $J_{n},J_{\theta},J_{\phi}$ transform under rotation
as vector operators due to their commutation relations\cite{sakurai},

\begin{eqnarray}
e^{-itJ_{n}}J_{\theta}e^{itJ_{n}} & = & J_{\theta}\cos t+J_{\phi}\sin t,\\
e^{-itJ_{n}}J_{\phi}e^{itJ_{n}} & = & -J_{\theta}\sin t+J_{\phi}\cos t.
\end{eqnarray}

Proceeding to calculate the components of the adiabatic connection
we get

\begin{eqnarray}
A_{\theta} & = & -\lim_{T\rightarrow\infty}\frac{B\mu}{2T}\text{\ensuremath{\int}}_{-T}^{T}dt\,\int_{0}^{t}ds\,e^{-isB\mu J_{n}}\,J_{\theta}\,e^{isB\mu J_{n}}\nonumber \\
 & = & -\lim_{T\rightarrow\infty}\frac{B\mu}{2T}\text{\ensuremath{\int}}_{-T}^{T}dt\,\int_{0}^{t}ds\,\left(\cos(sB\mu)J_{\theta}+\sin(sB\mu)J_{\phi}\right)\nonumber \\
 & = & -\lim_{T\rightarrow\infty}\frac{1}{2T}\text{\ensuremath{\int}}_{-T}^{T}dt\,\left(\sin(tB\mu)J_{\theta}-\cos(tB\mu)J_{\phi}+J_{\phi}\right)\nonumber \\
 & = & -J_{\phi},\\
A_{\phi} & = & -\lim_{T\rightarrow\infty}\frac{\sin\theta}{2T}\text{\ensuremath{\int}}_{-T}^{T}dt\,\int_{0}^{t}ds\,e^{-isB\mu J_{n}}\,J_{\phi}\,e^{isB\mu J_{n}}\nonumber \\
 & = & \sin\theta J_{\theta},\\
A_{B} & = & 0.
\end{eqnarray}

The only non-vanishing component of the curvature is $F_{\theta\phi}$.
The elements of $F_{\theta\phi}$ are

\begin{eqnarray}
\partial_{\theta}A_{\phi} & = & -\mathbf{J}\cdot\left(\partial_{\theta}\mathbf{\widehat{\mathbf{\phi}}}\right)=-\mathbf{J}\cdot\left(\widehat{n}\sin\theta-\widehat{\theta}\cos\theta\right)\nonumber \\
 & = & -\sin\theta J_{n}+\cos\theta J_{\theta},\\
\partial_{\phi}A_{\theta} & = & \sin\theta J_{n}+\cos\theta J_{\theta},\\
\left[A_{\theta},A_{\phi}\right] & = & i\sin\theta J_{n}.
\end{eqnarray}
The above equations can be put together to give
\begin{equation}
F=-\frac{1}{2}\sin\theta J_{n}\:d\theta\wedge d\phi.\label{F1}
\end{equation}

The Berry curvature associated to $\left|l,m\right\rangle _{\widehat{n}}$
is

\[
_{\widehat{n}}\left\langle l,m\right|F\left|l,m\right\rangle _{\widehat{n}}=-\frac{m}{2}\sin\theta\:d\theta\wedge d\phi.
\]

To give a concrete realization, let us consider the case $l=\frac{1}{2}$.
In this case, we can represent the angular momentum operator by the
Pauli matrices $J_{i}=\frac{1}{2}\sigma_{i}$. The representation
in spherical coordinates of the SU(2) algebra using the Pauli matrices
takes the form

\begin{align}
\mathbf{\overrightarrow{\sigma}\cdot\widehat{n}}=\sigma_{\widehat{n}} & =\left(\begin{array}{cc}
\cos(\theta) & e^{-i\phi}\sin(\theta)\\
e^{i\phi}\sin(\theta) & -\cos(\theta)
\end{array}\right),\nonumber \\
\sigma_{\theta} & =\left(\begin{array}{cc}
-\sin(\theta) & -e^{-i\phi}\cos(\theta)\\
e^{i\phi}\cos(\theta) & \sin(\theta)
\end{array}\right),\nonumber \\
\sigma_{\phi} & =\left(\begin{array}{cc}
0 & -e^{-i\phi}\\
e^{i\phi} & 0
\end{array}\right).
\end{align}
 The Hamiltonian becomes

\begin{equation}
H=\frac{B\mu}{2}\sigma_{\widehat{n}}.\label{Hpauli}
\end{equation}

Considering the original orientation to be $\widehat{n}=\widehat{z}$,
the starting Hamiltonian reduces to $H=\frac{B\mu}{2}\sigma_{z}$
with known eigenvectors given by

\begin{align}
\left|\uparrow,\widehat{z}\right\rangle  & =\left(\begin{array}{c}
1\\
0
\end{array}\right),\quad\left|\downarrow,\widehat{z}\right\rangle =\left(\begin{array}{c}
0\\
1
\end{array}\right).\label{vec}
\end{align}

We want to find the eigenvectors of (\ref{Hpauli}) for arbitrary
$\theta$ and $\phi$ by a parallel transport of the vectors (\ref{vec}).
The simplest path to take from the north pole $(\theta=0)$ to the
desired coordinates $(\theta,\phi)$ is to move across a great circle
with a fixed $\phi$. Since $A_{\theta}$ is independent of $\theta$
(so the connection commutes along the chosen path), there is no need
for a path ordering for the parallel transport operator

\begin{align}
e^{i\int_{0}^{\theta}A_{\theta}d\theta} & =e^{-i\frac{\theta}{2}\sigma_{\phi}}=\cos(\frac{\theta}{2})-i\sigma_{\phi}\sin(\frac{\theta}{2})\nonumber \\
 & =\left(\begin{array}{cc}
\cos(\frac{\theta}{2}) & -e^{-i\phi}\sin(\frac{\theta}{2})\\
e^{i\phi}\sin(\frac{\theta}{2}) & \cos(\frac{\phi}{2})
\end{array}\right).\label{Atheta}
\end{align}

Letting the matrix (\ref{Atheta}) act on the vectors (\ref{vec})
we obtain

\begin{equation}
\left|\uparrow,\widehat{n}\right\rangle =\left(\begin{array}{c}
\cos(\frac{\theta}{2})\\
e^{i\phi}\sin(\frac{\phi}{2})
\end{array}\right),\quad\left|\downarrow,\widehat{n}\right\rangle =\left(\begin{array}{c}
-e^{-i\phi}\sin(\frac{\theta}{2})\\
\cos(\frac{\phi}{2})
\end{array}\right),\label{n}
\end{equation}
a result that is in agreement with the known eigenvectors of (\ref{Hpauli})
\cite{sakurai}.

We will now check the lemma (\ref{lemma}) for a specific closed
path. Let the path start at the north pole and then descend to the
equator through the great circle $\phi=0$. The path continues in
the equator by an angle $\Omega$ to finally return to the north pole
via the great circle $\phi=\Omega$. The angle $\Omega$ traveled
in the equator also corresponds to the solid angle encircled by the
loop. The parallel transport operators associated with these segments
of the loop are

\begin{align}
e^{i\int_{0}^{\frac{\pi}{2}}A_{\theta}d\theta}= & e^{-i\frac{\pi}{4}\sigma_{\phi}(\phi=0)}=\left(\begin{array}{cc}
\cos(\frac{\pi}{4}) & -\sin(\frac{\pi}{4})\\
\sin(\frac{\pi}{4}) & \cos(\frac{\pi}{4})
\end{array}\right),\nonumber \\
e^{i\int_{0}^{\Omega}A_{\theta}(\theta=\frac{\pi}{2})d\theta}= & e^{i\frac{\Omega}{2}\sigma_{\theta}(\theta=\frac{\pi}{2})}=\left(\begin{array}{cc}
e^{-i\frac{\Omega}{2}} & 0\\
0 & e^{i\frac{\Omega}{2}}
\end{array}\right),\nonumber \\
e^{i\int_{\frac{\pi}{2}}^{0}A_{\theta}(\phi=\Omega)d\theta}= & e^{i\frac{\pi}{4}\sigma_{\phi}(\phi=\Omega)}=\left(\begin{array}{cc}
\cos(\frac{\pi}{4}) & e^{-i\Omega}\sin(\frac{\pi}{4})\\
-e^{i\Omega}\sin(\frac{\pi}{4}) & \cos(\frac{\pi}{4})
\end{array}\right).
\end{align}
The composition of the three paths gives

\begin{align}
e^{i\frac{\pi}{4}\sigma_{\phi}(\phi=\Omega)}e^{i\frac{\Omega}{2}\sigma_{\theta}(\theta=\frac{\pi}{2})}e^{-i\frac{\pi}{4}\sigma_{\phi}(\phi=0)} & =\left(\begin{array}{cc}
e^{-i\frac{\Omega}{2}} & 0\\
0 & e^{i\frac{\Omega}{2}}
\end{array}\right)\nonumber \\
 & =\exp\left(\begin{array}{cc}
-i\frac{\Omega}{2} & 0\\
0 & i\frac{\Omega}{2}
\end{array}\right).\label{spinphase}
\end{align}

We can see that (\ref{spinphase}) is indeed diagonal and that the
phases obtained correspond to the Berry phases for the spin $\frac{1}{2}$
system\cite{Berry}.

\subsection{Sp (2, $\mathbb{R}$)}

For the second example let us study the Hamiltonian for a generalized
oscillator

\begin{equation}
H=\frac{1}{2}(Xq^{2}+Y(pq+qp)+Zp^{2}),\label{H}
\end{equation}
limited to the bound state situation $ZX-Y^{2}>0$. For this Hamiltonian
the parameters to vary are $X$, $Y$ and $Z$.

To reduce (\ref{H}) to the standard form of the harmonic
oscillator we can perform the following symplectic transformation

\begin{eqnarray}
P & = & \frac{p\left[\frac{Z}{X}\right]^{1/4}+q\left[\frac{X}{Z}\right]^{1/4}}{\sqrt{2}}\left[\frac{\sqrt{XZ}+Y}{\sqrt{XZ}-Y}\right]^{1/4},\nonumber \\
Q & = & =\frac{-p\left[\frac{Z}{X}\right]^{1/4}+q\left[\frac{X}{Z}\right]^{1/4}}{\sqrt{2}}\left[\frac{\sqrt{XZ}-Y}{\sqrt{XZ}+Y}\right]^{1/4}.\label{trans}
\end{eqnarray}
The inverse transformation is

\begin{eqnarray}
q & = & \frac{1}{\sqrt{2}}\left[\frac{Z}{X}\right]^{1/4}\left(Q\left[\frac{\sqrt{XZ}+Y}{\sqrt{XZ}-Y}\right]^{1/4}+P\left[\frac{\sqrt{XZ}-Y}{\sqrt{XZ}+Y}\right]^{1/4}\right),\nonumber \\
p & = & \frac{1}{\sqrt{2}}\left[\frac{X}{Z}\right]^{1/4}\left(-Q\left[\frac{\sqrt{XZ}+Y}{\sqrt{XZ}-Y}\right]^{1/4}+P\left[\frac{\sqrt{XZ}-Y}{\sqrt{XZ}+Y}\right]^{1/4}\right).\label{Inversep}
\end{eqnarray}

In terms of the new variables (\ref{trans}), the Hamiltonian (\ref{H})
reduces to

\begin{equation}
H=\frac{\omega}{2}(P^{2}+Q^{2}),\label{HPQ}
\end{equation}
where $\omega=\sqrt{ZX-Y^{2}}$ and $\left[Q,P\right]=i$. In view
of Eq. (\ref{HPQ}), the Hamiltonian has eigenvectors that obey

\begin{equation}
H\left|n\right\rangle =E_{n}\left|n\right\rangle ,
\end{equation}
with

\begin{equation}
E_{n}=\omega\left(n+\frac{1}{2}\right),
\end{equation}
where $n$ is a non-negative integer.

As the Hamiltonian (\ref{HPQ}) is just a harmonic oscillator, the
operators $Q$ and $P$ obey the following evolution equations \cite{sakurai}

\begin{eqnarray}
e^{-itH}Qe^{itH} & = & Q(t)=Q\cos(\omega t)-P\sin(\omega t),\nonumber \\
e^{-itH}Pe^{itH} & = & P(t)=Q\sin(\omega t)+P\cos(\omega t).\label{QPevol}
\end{eqnarray}

We now have all the necessary tools to calculate the components of
the adiabatic connection. The $A_{Y}$ component can be obtained
as follows

\begin{eqnarray}
A_{Y} & = & -\lim_{T\rightarrow\infty}\frac{1}{4\pi}\int_{-T}^{T}\,t\,dt\int_{0}^{1}\,ds\,e^{-itsH}(\partial_{Y}H)e^{itsH}\nonumber \\
 & = & -\lim_{T\rightarrow\infty}\frac{1}{4\pi}\int_{-T}^{T}\,t\,dt\int_{0}^{1}\,ds\,e^{-itsH}(pq+qp)e^{itsH}\nonumber \\
 & = & -\lim_{T\rightarrow\infty}\frac{1}{4\pi}\int_{-T}^{T}\,t\,dt\int_{0}^{1}\,ds\,e^{-itsH}\left(P^{2}\left[\tfrac{\sqrt{(XZ)}-Y}{\sqrt{(XZ)}+Y}\right]^{1/2}-Q^{2}\left[\tfrac{\sqrt{(XZ)}+Y}{\sqrt{(XZ)}-Y}\right]^{1/2}\right)e^{itsH}\nonumber \\
 & = & -\lim_{T\rightarrow\infty}\frac{1}{4\pi}\int_{-T}^{T}\,t\,dt\,\nonumber \\
 &  & \times\int_{0}^{1}[\left(Q^{2}\sin^{2}(\omega st)+P^{2}\cos^{2}(\omega st)+\tfrac{1}{2}(QP+PQ)\sin(2\omega st)\right)\left[\tfrac{\sqrt{(XZ)}-Y}{\sqrt{(XZ)}+Y}\right]^{1/2}\nonumber \\
 &  & -\left(Q^{2}\cos^{2}(\omega st)+P^{2}\sin^{2}(\omega st)-\tfrac{1}{2}(QP+PQ)\sin(2\omega st)\right)\left[\tfrac{\sqrt{(XZ)}+Y}{\sqrt{(XZ)}-Y}\right]^{1/2}]\nonumber \\
 & = & \tfrac{\sqrt{XZ}}{8\omega^{2}}(QP+PQ)=\tfrac{\sqrt{XZ}}{8\omega^{2}}\left(\left[\frac{X}{Z}\right]{}^{\frac{1}{2}}q^{2}-\left[\frac{Z}{X}\right]^{\frac{1}{2}}p^{2}\right).\label{AY}
\end{eqnarray}

A similar procedure gives

\begin{eqnarray}
A_{Z} & = & -\frac{1}{16\omega^{2}}\left[\frac{X}{Z}\right]{}^{\frac{1}{2}}(2Q^{2}+Y(PQ+QP)-2P^{2})\nonumber \\
 & = & \frac{Y}{16\omega^{2}}\left[\frac{X}{Z}\right]{}^{\frac{1}{2}}\left(\left[\frac{Z}{X}\right]^{\frac{1}{2}}p^{2}-\left[\frac{X}{Z}\right]{}^{\frac{1}{2}}q^{2}\right)+\frac{Y^{2}}{8\omega^{3}}\left(\frac{X}{Z}q^{2}+p^{2}\right)\nonumber \\
 &  & +\frac{YX}{\omega^{2}}(qp+pq),\label{AZ}\\
A_{X} & = & -\frac{1}{16\omega^{2}}\left[\frac{Z}{X}\right]^{\frac{1}{2}}(2Q^{2}-Y(PQ+QP)-2P^{2})\nonumber \\
 & = & \frac{Y}{16\omega^{2}}\left[\frac{Z}{X}\right]^{\frac{1}{2}}\left(\left[\frac{Z}{X}\right]^{\frac{1}{2}}p^{2}-\left[\frac{X}{Z}\right]{}^{\frac{1}{2}}q^{2}\right)-\frac{Y^{2}}{8\omega^{3}}\left(q^{2}+\frac{Z}{X}p^{2}\right)\nonumber \\
 &  & -\frac{YZ}{\omega^{2}}(qp+pq).\label{AX}
\end{eqnarray}

The curvature can be computed to be

\begin{equation}
F=\frac{H}{8\omega^{3}}[X\,dY\wedge dZ+Z\,dX\wedge dY+Y\,dZ\wedge dX].\label{curvoscillator}
\end{equation}

The curvature is diagonal in the base of the eigenvectors of
$H$, with the non-vanishing entries given by

\begin{equation}
\left\langle n\right|F\left|n\right\rangle =\frac{n+1/2}{8\omega^{3}}[X\,dY\wedge dZ+Z\,dX\wedge dY+Y\,dZ\wedge dX],\label{berryoscillator}
\end{equation}
and they coincide with the Berry curvature for the nth eigenvector
for the generalized oscillator\cite{Berry}.

From a group theoretical point of view, consider the Lie algebra of
the symplectic group $sp(2,\mathbb{R})$

\begin{eqnarray}
[K_{1},K_{2}]=-iK_{0}, &  & [K_{0},K_{1}]=iK_{2},\nonumber \\
 & [K_{2},K_{0}]=iK_{1}.
\end{eqnarray}
This algebra has a representation in terms of the canonical pair of
operators $Q$ and $P$ given by

\begin{eqnarray}
K_{1}=\frac{1}{4}\left(Q^{2}-P^{2}\right), &  & K_{2},=-\frac{1}{4}\left(QP+PQ\right),\nonumber \\
 & K_{0}=\frac{1}{4}\left(Q^{2}+P^{2}\right).
\end{eqnarray}

We can see that the components of the adiabatic connection are a linear combination of $K_{1}$ and $K_{2}$. On the other hand, the
curvature is a function only of the generator of rotations $K_{0}$.
We refer to \cite{unitaroscillator} for a group theoretical analysis
of the relation between  $sp(2,\mathbb{R})$ and the generalized oscillator.

\section{Conclusions}

We used first-order Van Vleck-Primas perturbation theory to define
a parallel transport operator $Pe^{i\int_{\gamma}\mathcal{A}}$ that
relates the eigenvectors of a Hamiltonian $H(\lambda)$ at two different
points of parameter space. This allowed us to give two integral formulas for $\mathcal{A}$. Moreover, we showed that $\mathcal{A}$ can be
understood as a non-Abelian connection 1-form since it behaves correctly under non-Abelian gauge transformations, and it produces a parallel transport of the eigenbasis of the Hamiltonian. With the help of the
non-Abelian Stokes' theorem, we explicitly show that the holonomy of $\mathcal{A}$
is a diagonal operator whose non-vanishing entries are the Berry geometric phase of the respective eigenvectors.

An interesting property of the adiabatic connection is that it arises as an average of a Maurer-Cartan 1-form. This results in a connection with non-trivial holonomy even though the  Maurer-Cartan 1-form has zero curvature. The situation is reminiscent of the classical Hannay angles where a non-trivial parallel transport is defined by an average of a trivial connection \cite{Montgomery}. We hope to explore this possible relationship in the future.

From a practical point of view, the formulas given for the adiabatic connection are operational in nature. By not involving an infinite sum, they might prove to be easier to handle in calculations for transitionless driving in systems with an infinite number of eigenstates, as exemplified in our calculation for the generalized oscillator.

\appendix
\section{ Curvature}

The components of the Yang-Mills curvature are given by

\begin{equation}
F_{\mu\nu}=\partial_{\mu}A_{\upsilon}-\partial_{\upsilon}A_{\mu}-i\left[A_{\mu},A_{\upsilon}\right].\label{curvatureuv-1}
\end{equation}

We will show by direct computation that the curvature of the adiabatic
connection is diagonal on the original set of vectors and the
elements of the diagonal correspond to the Berry curvature. 

\subsubsection*{Diagonal Elements}

We can evaluate the diagonal elements of the commutator as

\begin{eqnarray*}
\left\langle n\right|\left[A_{\mu},A_{\upsilon}\right]\left|n\right\rangle  & = & \left\langle n\right|A_{\mu}A_{\upsilon}\left|n\right\rangle -\left\langle n\right|A_{\upsilon}A_{\mu}\left|n\right\rangle \\
 & = & \sum_{n\neq n'}\left\{ \left\langle n\right|A_{\mu}\left|n'\right\rangle \left\langle n'\right|A_{\upsilon}\left|n\right\rangle -\left\langle n\right|A_{\upsilon}\left|n'\right\rangle \left\langle n'\right|A_{\mu}\left|n\right\rangle \right\} \\
 & = & \sum_{n\neq n'}\left\{ \frac{\left\langle n\right|\partial_{\mu}H\left|n'\right\rangle \left\langle n'\right|\partial_{\nu}H\left|n\right\rangle }{\left(E_{n}-E_{n'}\right)^{2}}-\frac{\left\langle n\right|\partial_{\nu}H\left|n'\right\rangle \left\langle n'\right|\partial_{\mu}H\left|n\right\rangle }{\left(E_{n}-E_{n'}\right)^{2}}\right\} ,
\end{eqnarray*}
where we have used Eq. (\ref{eq:per}) and the parallel transport
condition (\ref{parallelcondition}). The derivative terms can be
computed as follows

\begin{eqnarray*}
\left\langle n\right|\partial_{\mu}A_{\upsilon}\left|n\right\rangle  & = & \partial_{\mu}\left\langle n\right|A_{\upsilon}\left|n\right\rangle (=0)\\
 &  & -\left\langle \partial_{\mu}n\right|A_{\upsilon}\left|n\right\rangle -\left\langle n\right|\partial_{\mu}A_{\upsilon}\left|\partial_{\mu}n\right\rangle \\
 & = & i\left\langle n\right|A_{\mu}A_{\upsilon}\left|n\right\rangle -i\left\langle n\right|A_{\upsilon}A_{\mu}\left|n\right\rangle \\
 & = & i\left\langle n\right|\left[A_{\mu},A_{\upsilon}\right]\left|n\right\rangle .
\end{eqnarray*}
Similarly

\[
\left\langle n\right|\partial_{\upsilon}A_{\mu}\left|n\right\rangle =-i\left\langle n\right|\left[A_{\mu},A_{\upsilon}\right]\left|n\right\rangle .
\]
Thus, 

\begin{eqnarray*}
\left\langle n\right|F_{\mu\nu}\left|n\right\rangle  & = & i\left\langle n\right|\left[A_{\mu},A_{\upsilon}\right]\left|n\right\rangle \\
 & = & i\sum_{n\neq n'}\left\{ \frac{\left\langle n\right|\partial_{\mu}H\left|n'\right\rangle \left\langle n'\right|\partial_{\nu}H\left|n\right\rangle }{\left(E_{n}-E_{n'}\right)^{2}}-\frac{\left\langle n\right|\partial_{\nu}H\left|n'\right\rangle \left\langle n'\right|\partial_{\mu}H\left|n\right\rangle }{\left(E_{n}-E_{n'}\right)^{2}}\right\} .
\end{eqnarray*}

\subsubsection*{Off-Diagonal Elements}

The off-diagonal elements of the commutator are

\begin{eqnarray*}
\left\langle m\right|\left[A_{\mu},A_{\upsilon}\right]\left|n\right\rangle  & = & -\sum_{n\neq n'}\left\{ \frac{\left\langle m\right|\partial_{\mu}H\left|n'\right\rangle \left\langle n'\right|\partial_{\nu}H\left|n\right\rangle }{\left(E_{m}-E_{n'}\right)\left(E_{n'}-E_{n}\right)}-\frac{\left\langle m\right|\partial_{\nu}H\left|n'\right\rangle \left\langle n'\right|\partial_{\mu}H\left|n\right\rangle }{\left(E_{m}-E_{n'}\right)\left(E_{n'}-E_{n}\right)}\right\} .
\end{eqnarray*}
For the derivative terms, we have to take into account that this time the terms $\partial_{\mu}\left\langle m\right|A_{\upsilon}\left|n\right\rangle $
and $\partial_{\nu}\left\langle m\right|A_{\mu}\left|n\right\rangle $
do not vanish. We can write for the derivative terms the following

\begin{eqnarray*}
\left\langle m\right|\partial_{\mu}A_{\upsilon}\left|n\right\rangle  & = & \partial_{\mu}\left\langle m\right|A_{\upsilon}\left|n\right\rangle +i\left\langle m\right|\left[A_{\mu},A_{\upsilon}\right]\left|n\right\rangle ,\\
\left\langle m\right|\partial_{\mu}A_{\upsilon}\left|n\right\rangle  & = & \partial_{\nu}\left\langle m\right|A_{\mu}\left|n\right\rangle -i\left\langle m\right|\left[A_{\mu},A_{\upsilon}\right]\left|n\right\rangle .
\end{eqnarray*}
So far the off-diagonal elements of the curvature are 

\[
\left\langle m\right|F_{\mu\nu}\left|n\right\rangle =\partial_{\mu}\left\langle m\right|A_{\upsilon}\left|n\right\rangle -\partial_{\nu}\left\langle m\right|A_{\mu}\left|n\right\rangle +i\left\langle m\right|\left[A_{\mu},A_{\upsilon}\right]\left|n\right\rangle .
\]

Noting the identity

\[
\frac{1}{\left(E_{m}-E_{n}\right)\left(E_{m}-E_{n'}\right)}+\frac{1}{\left(E_{m}-E_{n}\right)\left(E_{n'}-E_{n}\right)}=\frac{1}{\left(E_{m}-E_{n'}\right)\left(E_{n'}-E_{n}\right)},
\]
we can write the following

\begin{eqnarray*}
\partial_{\mu}\left\langle m\right|A_{\upsilon}\left|n\right\rangle -\partial_{\nu}\left\langle m\right|A_{\mu}\left|n\right\rangle  & = & \frac{1}{E_{m}-E_{n}}\left(\partial_{\mu}\left\langle m\right|\partial_{\nu}H\left|n\right\rangle -\partial_{\nu}\left\langle m\right|\partial_{\mu}H\left|n\right\rangle \right)\\
 & = & \frac{1}{E_{m}-E_{n}}\left\{ \left\langle \partial_{\mu}m\right|\partial_{\nu}H\left|n\right\rangle +\left\langle m\right|\partial_{\nu}H\left|\partial_{\mu}n\right\rangle \right.\\
 &  & \left.-\left\langle \partial_{\nu}m\right|\partial_{\mu}H\left|n\right\rangle -\left\langle m\right|\partial_{\mu}H\left|\partial_{\nu}n\right\rangle \right\} \\
 & = & -\frac{i}{E_{m}-E_{n}}\left\{ \sum_{m\neq n'}\frac{\left\langle m\right|\partial_{\mu}H\left|n'\right\rangle \left\langle n'\right|\partial_{\nu}H\left|n\right\rangle }{\left(E_{m}-E_{n'}\right)}\right.\\
 &  & -\sum_{n\neq n'}\frac{\left\langle m\right|\partial_{\nu}H\left|n'\right\rangle \left\langle n'\right|\partial_{\mu}H\left|n\right\rangle }{\left(E_{n'}-E_{n}\right)}\\
 &  & -\sum_{m\neq n'}\frac{\left\langle m\right|\partial_{\nu}H\left|n'\right\rangle \left\langle n'\right|\partial_{\mu}H\left|n\right\rangle }{\left(E_{m}-E_{n'}\right)}\\
 &  & \left.+\sum_{n\neq n'}\frac{\left\langle m\right|\partial_{\mu}H\left|n'\right\rangle \left\langle n'\right|\partial_{\nu}H\left|n\right\rangle }{\left(E_{n'}-E_{n}\right)}\right\} \\
 & = & -i\sum_{n\neq n'}\left\{ \frac{\left\langle m\right|\partial_{\mu}H\left|n'\right\rangle \left\langle n'\right|\partial_{\nu}H\left|n\right\rangle }{\left(E_{m}-E_{n'}\right)\left(E_{n'}-E_{n}\right)}\right.\\
 &  & \left.-\frac{\left\langle m\right|\partial_{\nu}H\left|n'\right\rangle \left\langle n'\right|\partial_{\mu}H\left|n\right\rangle }{\left(E_{m}-E_{n'}\right)\left(E_{n'}-E_{n}\right)}\right\} \\
 & = & i\left\langle m\right|\left[A_{\mu},A_{\upsilon}\right]\left|n\right\rangle 
\end{eqnarray*}

Hence, we finally have

\[
\left\langle m\right|F_{\mu\nu}\left|n\right\rangle =0.
\]

\section{ Non-Abelian Stokes' Theorem}

For an Abelian connection 1-form, the Stokes's theorem states that

\begin{equation}
\oint_{\gamma}\mathcal{A}=\int_{S}F,\label{abelian}
\end{equation}
where $F=d\mathcal{A}$ is the Abelian curvature 2-form. The underlying idea of the Abelian Stokes theorem for 1-forms is exemplified in Figure
1. The initial closed path is decomposed into smaller loops, and each
loop has associated an area element, allowing for a transition from
original line integral to a surface integral in the limit where the
loops get infinitesimally small. Let us remark that, as long as the
orientation is preserved, each loop can be taken independently to
the others due to the Abelian character of the connection.

\begin{figure}[H]
\includegraphics[scale=0.35]{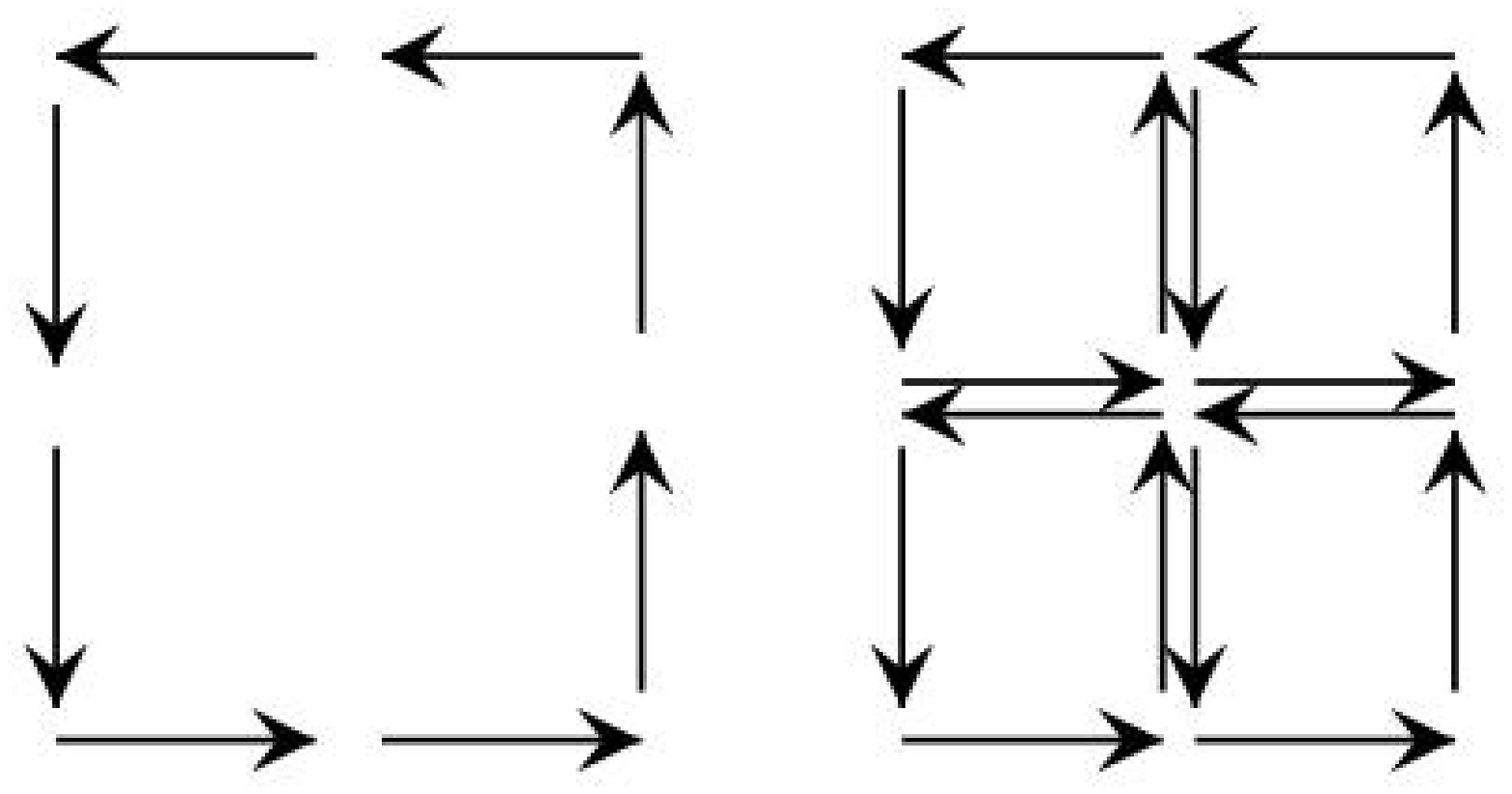}
\caption{Decomposition of a loop for the Stokes theorem}
\end{figure}

A result like (\ref{abelian}) is impossible for non-Abelian 1-forms
\cite{stokes 5}. However, there is a generalization of the Stokes'
theorem for ordered exponential of a non-Abelian 1-form $\mathcal{A}(\lambda)=A_{\mu}d\lambda^{\mu}$.
If $\mathcal{A}$ and the curve $\gamma=\partial S$ are continuously
differentiable, then the non-Abelian Stokes' theorem (NAST) states
that the path ordered exponential of $\mathcal{A}$ can be replaced
by a surface ordered one

\begin{equation}
Pe^{i\oint_{\gamma}\mathcal{A}}=\mathcal{P}e^{i\int_{S}\mathcal{F}},\label{nonabelian-1}
\end{equation}
where $\mathcal{F}$ is known as the ``twisted'' or \textquotedblleft path-dependent\textquotedblright{}
curvature 2-form and $\mathcal{P}$ represents a suitable surface
ordering. Here we will explain the meaning of all the terms appearing
in the right hand side of Eq. (\ref{nonabelian-1}), and we will give
an intuitive justification of the NAST. For rigorous proof of the
theorem see \cite{stokes}, our exposition is based on \cite{stokes2}.

The parallel transport operator along a curve starting at $\lambda'$
and ending at $\lambda''$ is defined by

\begin{equation}
U(\lambda'',\lambda')=P\,\exp\left[\int_{\lambda'}^{\lambda''}A_{\mu}(\lambda)d\lambda^{\mu}\right],\label{parallelU}
\end{equation}
where the parametrization of the curve has been left implicit. Note
that the inverse operator of $U$ correspond to traveling the path
backward

\begin{equation}
U^{-1}(\lambda'',\lambda')=U(\lambda',\lambda'').
\end{equation}
The components of twisted curvature are defined by

\begin{equation}
\mathcal{F}_{\mu\nu}(\lambda)\overset{def}{=}U^{-1}(\lambda,\lambda_{0})F_{\mu\nu}(\lambda)U(\lambda,\lambda_{0}).\label{twistedF}
\end{equation}
The geometrical meaning of (\ref{twistedF}) is the following: Eq.
(\ref{twistedF}) represent the holonomy of lassos like the one depicted
in Figure 2. The lasso starts with $U(\lambda,\lambda_{0})$ joining
the base point with a given point $\lambda$. The next operation is
an infinitesimal loop of area 2-form $\varepsilon^{2}$. The lasso
returns to the base point via $U^{-1}(\lambda,0)$. The holonomy of
this closed path is

\begin{equation}
U^{-1}(\lambda,\lambda_{0})e^{i\varepsilon^{2}F(\lambda)}U(\lambda,\lambda_{0})=e^{i\varepsilon^{2}\mathcal{F}}.\label{lassos}
\end{equation}
Given a parametrization on the surface $\lambda=\lambda(x_{i},x_{j})$,
with $\lambda_{0}$ given by $(0,0)$, the lasso can be labeled by
the coordinates of the loop 
\begin{equation}
U_{ij}^{-1}e^{i\varepsilon_{ij}^{2}Fij}U_{ij}.\label{lassos2}
\end{equation}

\begin{figure}[H]
\includegraphics[scale=0.35]{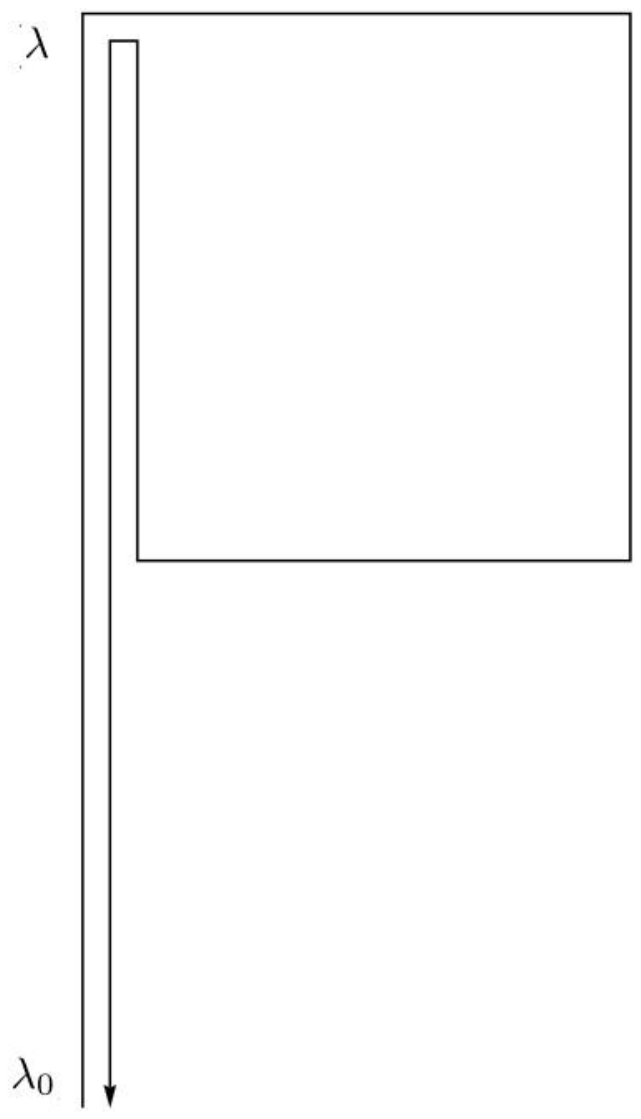}
\caption{Single lasso}
\end{figure}

The NAST states that (a) the surface $S$ can be filled by the union of
lassos in the form given by (\ref{lassos}), and (b) there exists a proper order to multiply the infinitesimal lassos around different
point such that the result is equal to the path ordered
exponential.

\begin{equation}
\mathcal{P}e^{i\int_{S}\mathcal{F}}=\lim_{N\rightarrow\infty}\mathcal{P}\prod_{i,j}^{N}U_{i,j}^{-1}\,e^{i\varepsilon_{i,j}^{2}F_{i,j}}\,U_{i,j}=Pe^{i\oint_{\gamma}\mathcal{A}},\label{PF-1}
\end{equation}
Figure 3 illustrates a way to decompose into two lassos the rectangle
with coordinates on the surface given by $\left\{ (0,0),\,(0,2\varepsilon),\,(\varepsilon,2\varepsilon),\,(\varepsilon,0)\right\} $.
Note that the order of the lassos is crucial for the cancellation
of the internal lines. If the lasso of the bottom part were to be
done first, then the original path ordered exponential would not be
recovered. In this case, the surface ordering multiplication of the
lassos reads

\begin{eqnarray}
\mathcal{P}\prod_{i,j} & U_{ij}^{-1}e^{i\varepsilon_{ij}^{2}Fij}U_{ij}= & \left(U^{-1}(0,\varepsilon)e^{i\varepsilon^{2}F(0,\varepsilon)}U(0,\varepsilon)\right)\times\nonumber \\
 &  & \left(U^{-1}(0,2\varepsilon)e^{i\varepsilon^{2}F(0,2\varepsilon)}U(0,2\varepsilon)\right).
\end{eqnarray}

\begin{figure}[H]
\includegraphics[scale=0.3]{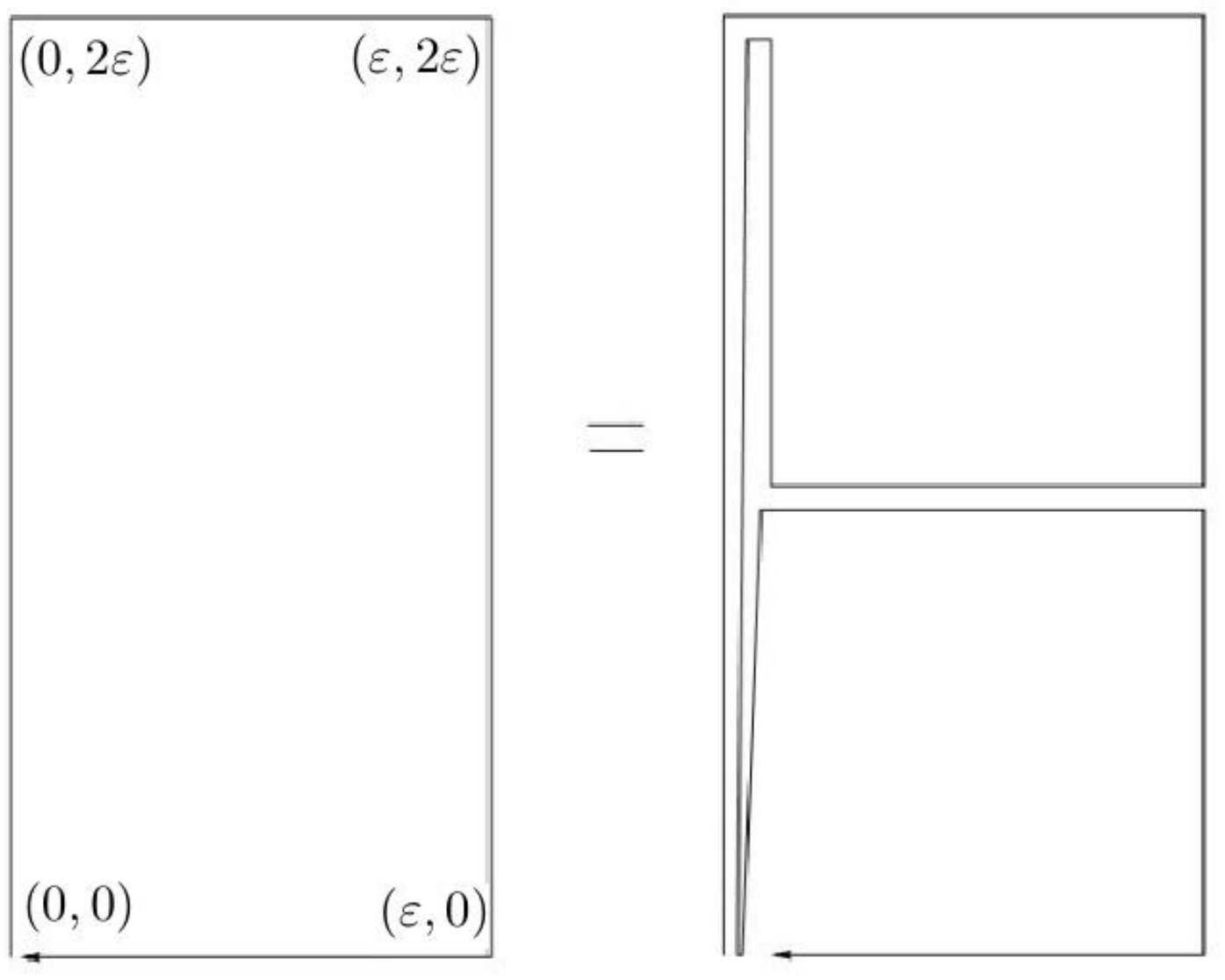}
\caption{Decomposition of a loop into two lassos}
\end{figure}

Let us make two final remarks. First, the value of $\mathcal{P}e^{i\int_{S}\mathcal{F}}$
is invariant under continuous changes of the surface as long as $\gamma=\partial S$
is maintained\cite{stokes 4}. Lastly, there is a rigorous way to
give a precise rule for the path ordering $\mathcal{P}$ for general
surfaces, but, for our purposes here, we do not need it due to the
properties of the adiabatic connection as explained in section 2.3.

\end{document}